\documentclass[10pt]{article}
\usepackage{graphicx}
\begin{document}
\title{THE SEARCH FOR POSSIBLE MESOLENSING OF COSMIC GAMMA-RAY BURSTS.
DOUBLE AND TRIPLE BURSTS IN BATSE CATALOGUE}
\author{O.S. Ougolnikov}
\date{}
\maketitle

\begin{center}
Astro-Space Center Of Lebedev's Physical Institute, 117997, Profsoyuznaya 84/32, Moscow, Russia\\
ugol@tanatos.asc.rssi.ru
\end{center}
The work is devoted to the search for possible effect of gravitational lensing
on the objects like globular stellar clusters (mesolensing) on the lightcurves
and spectra of large number of gamma-ray bursts from BATSE catalogue. The
general properties and different types of such event were considered. As the
result 11 possibly mesolensed gamma-ray bursts were found. The propeties of
possible mesolenses are invesigated.

\section{Introduction}

The gravitational lensing event predicted by the General Theory of Relativity
were discovered in 1979 [1] for the distant quasar, which image was duplicated
by massive elliptical galaxy situated between quasar and observer. Since that
time many lensed extragalactic objects (quasars, AGN) were found.

The last years of XX century the hypothesis of extragalactic origin of
cosmic gamma-ray bursts [2] was confirmed (at least for sufficient part of
gamma-ray bursts) by X-, optical and radio afterglows discovering (see [3]
for example). But in this case we should observe the gravitational lensing
effect for these objects too, as it was firstly reported in [4]. But a number
of papers dedicated to the search of lensed gamma-ray bursts [5,6] had a
negative result.

In this work we shall consider the so-called mesolensing event, when the role
of gravitational lens is played by the body like globular stellar cluster in
far galaxy or other body with mass about $10^6 M_{\odot}$. The probability of
observation of such type of lensing can be quite large due to the sufficient
total mass of such objects in the Universe [7,8], especially at large values
of $z$ [9]. Mesolensing can be very often phenomenon for extragalactic sources, and
possibility of very high amount of magnification of the image makes it one of
possible explanation of observing correlation of quasars and nearby galaxies
[10]. The primary goal of this work is to find gamma-ray bursts having the
mesolensed properties.

\section{Observational Properties of Mesolensed\\ Gamma-Ray Bursts}

The most simple case of gravitational lensing by a point mass, when the
light deflecting angle is back proportional to the distance of light path from
the lens, was considered many times [11]. The scheme of this phenomenon is shown
on Fig.1a. In this case an observer $O$ will see two images of the source $S$
at different sides from the lens $L$. Angular distances of images from the lens
$\theta_{1,2}$ will be equal to
\begin{equation}
\label{1}
\theta_{1,2}=\frac{\alpha}{2}\pm\sqrt{\left(\frac{\alpha}{2}\right)^2+\theta_0^2}\mbox{,}
\end{equation}\\
where $\alpha$ is the angle between source and lens directions and $\theta_0$
is angular Enstein radius defined by the formula
\begin{equation}
\label{2}
\theta_0=\sqrt{\frac{4GML_{LS}}{c^2L_{OS}L_{OL}}}\mbox{,}
\end{equation}\\
where $M$ is the mass of the lens. However in the case of gamma-ray bursts we
cannot resolve two different lensed images with separation about 1${''}$, but
we can detect another effect related with difference of time of light
propagation along trajectories 1 and 2. This effect appears for two reasons:
geometrical difference of path lengths and relativistic time delay near the
mass $L$ (Shapiro effect). This effects are adding to each other and finally
burst signals by path 1 will reach the observer earlier than by path 2 and
total time delay between moments of registration will be equal to
\begin{equation}
\label{3}
\Delta t=\frac{2GM\left(1+z_L\right)}{c^3}
\left[\frac{\theta_1^2-\theta_2^2}{\theta_0^2}+\ln\frac{\theta_1^2}{\theta_2^2}\right]\mbox{,}
\end{equation}\\
where $z_L$ is the lens redshift. In the case of gravitational lensing the
gamma-ray burst will have two lightcurve components with similar profiles
(differing only by a factor of constant) and similar spectra. Magnification of
two images will be different and image 1 will be $K$ times brighter than image
2, and brightness ratio can be expressed by the formula
\begin{equation}
\label{4}
K=\frac{\theta_1^2}{\theta_2^2}\mbox{.}
\end{equation}\\
Comparing the formulae (\ref{3}) and (\ref{4}) we obtain relation of two
observational parameters $\Delta t$ and $K$:
\begin{equation}
\label{5}
\Delta t=\frac{2GM\left(1+z_L\right)}{c^3}\left[\frac{K-1}{\sqrt{K}}+\ln{K}\right]\mbox{.}
\end{equation}\\
Here we can see that for given $\Delta t$ and $K$ from observations one may
obtain mass $M$ if we can assume small $z_L$ and small size of the lens. We
can also see that for $K>1$ $\Delta t>0$ and in the case of point mass lens
bright image will be registered earlier than faint one.

\begin{figure}[t]
\centering
\resizebox{1.0\textwidth}{!}{\includegraphics{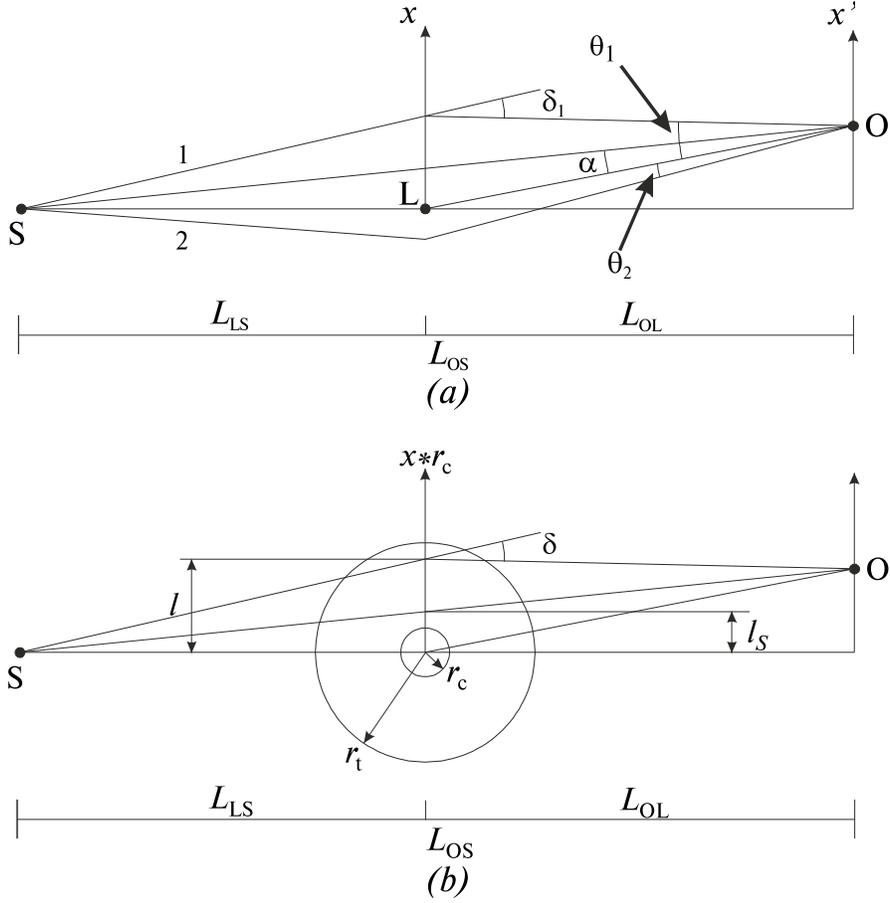}}
\caption{Gravitational lensing effect for point mass lens $(a)$ and for
globular cluster $(b)$.}
\end{figure}

As we can see from (\ref{5}) time delay $\Delta t$ is proportional to the
lens mass $M$ and type of lens will define type of lensing. In the case of
macrolensing when the galaxy or galaxy cluster with mass about 
$10^{11}-10^{13}M_{\odot}$ plays the role of lens the time delay will be
equal to several months or years and in the same case we should register two
different gamma-ray bursts from the one region on the sky. The search for this
type of lensing was conducted in [5,6], but the result was negative.

In the case of microlensing on a singular star the time delay, vice versa, will
be very small, about $10^{-5} sec$ and such duplication of burst lightcurve is
very difficult to detect. Finally, in the case of mesolensing the amount of
$\Delta t$ will be equal to several seconds what is comparable with the burst
duration, and in this case we shall see one burst with duplication of its
lightcurve on two similar (by a factor of constant) components with equal
spectra.

But possible gravitational lens may be extended object with definite size and 
in this case formulae above will not be correct. To estimate the range of
valiability of point mass assumption we compare the external radius of
globular cluster $r_t$ and spatial Einstein radius $\theta_0 L_{OL}$. This
leads us to the limitation to the lens distance:
\begin{equation}
\label{6}
L_0\geq\frac{r_t^2 c^2 L_{OS}}{4GML_{LS}}\geq\frac{r_t^2 c^2}{4GM}\mbox{.}
\end{equation}\\
If we take values $r_t=30pc$ and $M=5*10^5M_{\odot}$ that are typical for
globular clusters [12], we shall obtain $L_0$ about $10^{10} pc$ which is
comparable with the size of the Universe. This means that for all smaller
distances of mesolenses we have to take into account their spatial mass
disribution and light rays going inside of cluster. We shall call this case
as ``near mesolensing" and the case of point mass lens as ``far mesolensing".

The problem of mesolensing with account of inner structure of globular
clusters was investigated in [13] and developed in [10]. The scheme of
near mesolensing is shown on Fig.1b. The density distribution for globular
cluster is defined by a King law [14]:
\begin{eqnarray}
\label{7}
\rho\left(p\right)=\rho_0\left(1+\frac{p^2}{r_c^2}\right)^{-\frac{3}{2}},p<r_t \nonumber  \\
\rho\left(p\right)=0, p\geq r_t\mbox{,}
\end{eqnarray}\\
where $p$ is the distance from the cluster center, $\rho_0$ is the central
density of the cluster and $r_c$ is the core radius of the cluster. Following
[13] and [10], we express the distances $l$ (from ray path to the lens center),
$l_s$ (from lens center to the ``source-observer" line), and $r_t$ in the units
of $r_c$:
\begin{equation}
\label{8}
x=\frac{l}{r_c}, y=\frac{l_s}{r_c}, x_t=\frac{r_t}{r_c}\mbox{.}
\end{equation}\\
In this unit system the deflecing angle $\delta$ will be expressed as
\begin{equation}
\label{9}
\delta\left(l\right)=\frac{r_c}{F_L}\alpha\left(x\right)\mbox{,}
\end{equation}\\
where $F_L$ (lens focal distance) is determined as
\begin{equation}
\label{10}
F_L=\frac{c^2}{8\pi G\rho_0 r_c}\mbox{,}
\end{equation}\\
and function $\alpha(x)$ is following:
\begin{eqnarray}
\label{11}
\alpha\left(x\right)=\frac{ln\left(1+x^2\right)}{x}-
4\frac{\sqrt{1+x^2}-1}{x\sqrt{1+x_t^2}}+\frac{x}{1+x_t^2}, x\leq x_t \nonumber \\
\alpha\left(x\right)=\alpha\left(x_t\right)\frac{x_t}{x}, x>x_t\mbox{.}
\end{eqnarray}\\
The condition of registration of the light ray by observer may be written as
\begin{equation}
\label{12}
\alpha\left(x\right)=\frac{x-y}{g}\mbox{,}
\end{equation}\\
where the spatial parameter $g$ is defined by the formula
\begin{equation}
\label{13}
g=\frac{L_{OL}L_{LS}}{L_{OS}F_L}\mbox{.}
\end{equation}\\
Given parameters $g$, $y$ and $x_t$, formulae (\ref{11}) and (\ref{12}) are
building an equation for $x$, each solution $x_i$ of this equation will
correspond to one source image an observer notices. The magnification of
this image can be expressed by the formula
\begin{equation}
\label{14}
I_i=\left|\frac{x_i}{y}\frac{dx_i}{dy}\right|\mbox{.}
\end{equation}\\
The time of registration of the $i$-image $t_i$ relatively to the time of
light propagation through the center of cluster $t_0$ will be equal to
\begin{equation}
\label{15}
t_i-t_0=\frac{8\pi G\rho_0 r_c^3\left(1+z_L\right)}{c^3}T\left(x_i\right)=C_0 T\left(x_i\right)\mbox{,}
\end{equation}\\
where the function $T(x)$ is expressed by the formula
\begin{equation}
\label{16}
T\left(x\right)=2x\alpha\left(x\right)-2\int_0^x\alpha\left(x'\right)dx'-\frac{x^2}{g}\mbox{.}
\end{equation}\\
Analysis of last formulae shows that at $g>1$ there are two caustics. Axial
caustic corresponds $y=0$ and this case is the analog to the "Einstein Ring"
for the point mass lensing, but now we shall have weak central image along with
bright ring. There is also the conic caustic at $|y|=y_c(g,x_t)$, when two
images merge to one and magnification of this image turns to infinity too.
The conic caustic is clearly visible on Fig.2. Axial and conic caustics merge
in the focal point of the lens.

\begin{figure}[t]
\centering
\resizebox{1.0\textwidth}{!}{\includegraphics{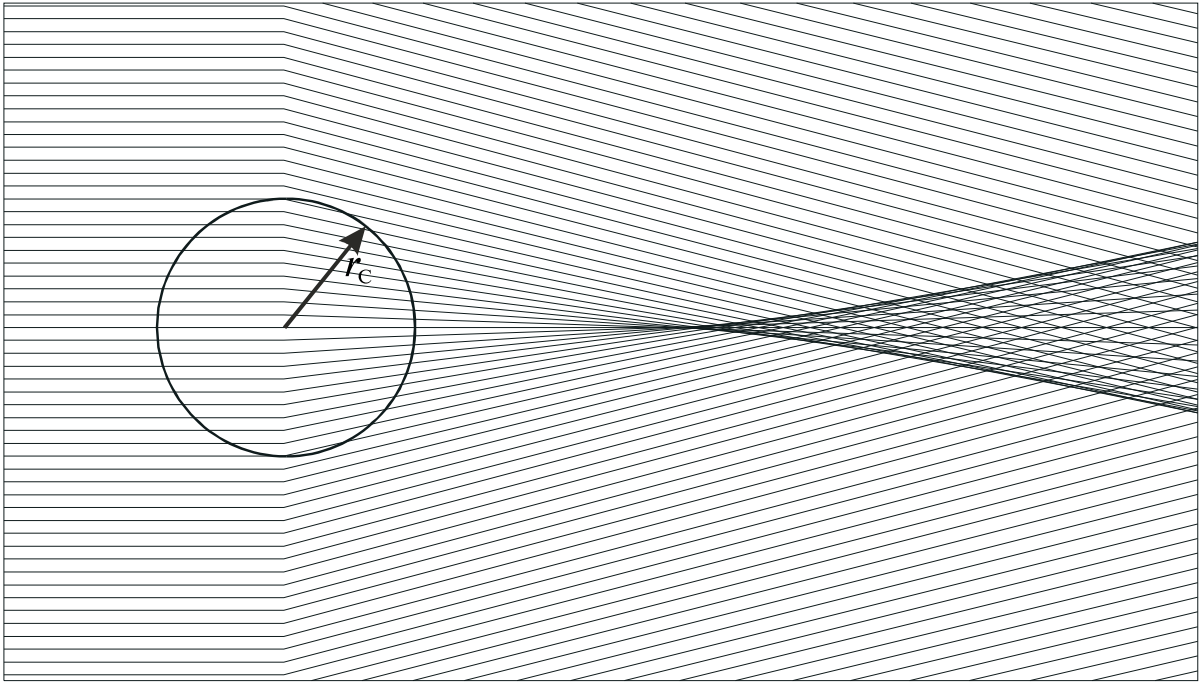}}
\caption{Light ray paths for near mesolensing, the conic caustic.}
\end{figure}

The number of images is equal to one outside the conic caustic but rise up to
three inside it ($|y|<y_c(g,x_t)$). Thus, in the case of near mesolensing of
cosmic gamma-ray bursts we can see double or triple lightcurve, but all
components should be similar (by the factor of constant) and have similar
spectra again. Numerical analysis of formulae (\ref{14}) and (\ref{15}) shows
that in most part of cases first (by the time of registration) component
will be brightest as it took place in the point-mass lensing case, but now
this rule can be broken, for example, near the conic caustic surface.

Far and near mesolensing properties discussed above shows that generally we
should search for the double and triple gamma-ray bursts with any combinations
of component brightness, but bursts with bright first lightcurve component
will be more probable candidates to the mesolensed ones.

\section{The search for possible mesolensed gamma-ray bursts}

The search for possible mesolensed bursts were conducted amongst 1512 events
registered by BATSE [15] during the whole CGRO work period from 1991 to 2000
and having MER format data with lightcurves with resolution 16 and 64 $ms$
and 16 channel spectra. At the first stage candidates with visually similar
two or three lightcurve components were selected and approximate amounts of
brightness ratio and time delay between them were estimated. At the second
stage the similarity of these lightcurve components was tested by
$\chi^2$-method with 0.003 significance level (this test is analogical to
the $3\sigma$-test) and the values of brightness ratio and time delay were
precisely calculated by best $\chi^2$-agreement method. Finally, third stage
of the test was $\chi^2$-comparison of the 15 channel spectra of the
components with the same significance level (channel 16 data with highest
energy with maximum error was ignored). The method is similar to that developed
in [16].

\subsection{Double gamma-ray bursts}

As the result of the search we have found 11 double gamma-ray bursts which
two lightcurve components had satisfied all the statistical tests discussed
above. Fig.3 shows lightcurves of all 11 candidates and Table 1 contains their
parameters: burst name $NB$, BATSE trigger number $Tr$, burst galactic
coordinates $l$ and $b$, their uncertainty $\Delta\epsilon$, brightness ratio
of two components $K$ and time delay between their registration moments
$\Delta t$. 6 of these candidates from 4 BATSE Catalogue [15], from
GRB 930430B to GRB 960617B  were firstly described in paper [17].

First of all we should emphasize on the fact that the brightest component
appeared to be first one for all 11 candidates although, unlike [17], it was
not required at the search procedure! This situation is difficult to imagine
in the case of occasional similarity of two burst components and it 
sufficiently increases the probability of real mesolensing for even though 
some of these bursts.

Assuming the possible lens to be point mass with not so large $z_L$ we can
estimate its mass by formula (\ref{5}) for our 11 far mesolensed candidates.
These values of mass are given in the last column of Table 1. For 6 described
in [17] candidates these values of mass differ from noted in [17] due to
taking Shapiro effect into account here. We can see that these values are
suitable for globular clusters or objects described in [7--9]. We should also
note that gamma-ray burst GRB 960617B  is situated at $4^{\circ}$
(about $2\Delta\epsilon$) from galaxy cluster AGC 1060 in Hydra.

\begin{table}[!h]
\begin{center}
\caption{Double bursts parameters.}
\begin{tabular}{|l|r|r|r|c|c|r|c|}
\hline
Burst & $Tr$ & $l$ & $b$ & $\Delta\epsilon$ & $K$ & $\Delta t$ & $M$ \\
\hline
 & & $^\circ$ & $^\circ$ & $^\circ$ & & $sec$ & $M_{\odot} $ \\
911006  &  871 & 266.7 & --65.5 & 3.4 & 2.498 &  4.560 & $ 2.5*10^5 $\\
920113  & 1297 & 139.9 &  +23.1 & 4.0 & 1.447 &  4.544 & $ 6.2*10^5 $\\
920728  & 1729 &  76.6 & --15.4 & 4.4 & 4.854 &  9.328 & $ 2.8*10^5 $\\
930430B & 2322 &  65.2 & --56.9 & 1.7 & 4.850 & 45.952 & $ 1.4*10^6 $\\
930602  & 2367 &  62.5 &  +31.0 & 2.5 & 8.986 &  9.696 & $ 2.0*10^5 $\\
931008C & 2569 & 353.9 & --15.7 & 4.5 & 1.571 &  7.744 & $ 8.7*10^5 $\\
960601  & 5483 & 119.8 &   +2.1 & 3.9 & 1.890 &  1.120 & $ 8.8*10^4 $\\
960615C & 5499 & 330.2 & --29.7 & 7.5 & 1.450 &  0.432 & $ 5.9*10^4 $\\
960617B & 5504 & 273.2 &  +24.5 & 2.2 & 4.562 &  3.920 & $ 1.2*10^5 $\\
000126A & 7968 & 111.8 & --26.3 & 4.5 & 1.101 & 11.968 & $ 6.3*10^6 $\\
000407  & 8066 & 291.5 & --10.4 & 2.2 & 2.119 &  1.952 & $ 1.3*10^5 $\\
\hline
\end{tabular}
\end{center}
\end{table}
					      
\begin{figure}[!h]
\centering
\resizebox{1.0\textwidth}{!}{\includegraphics{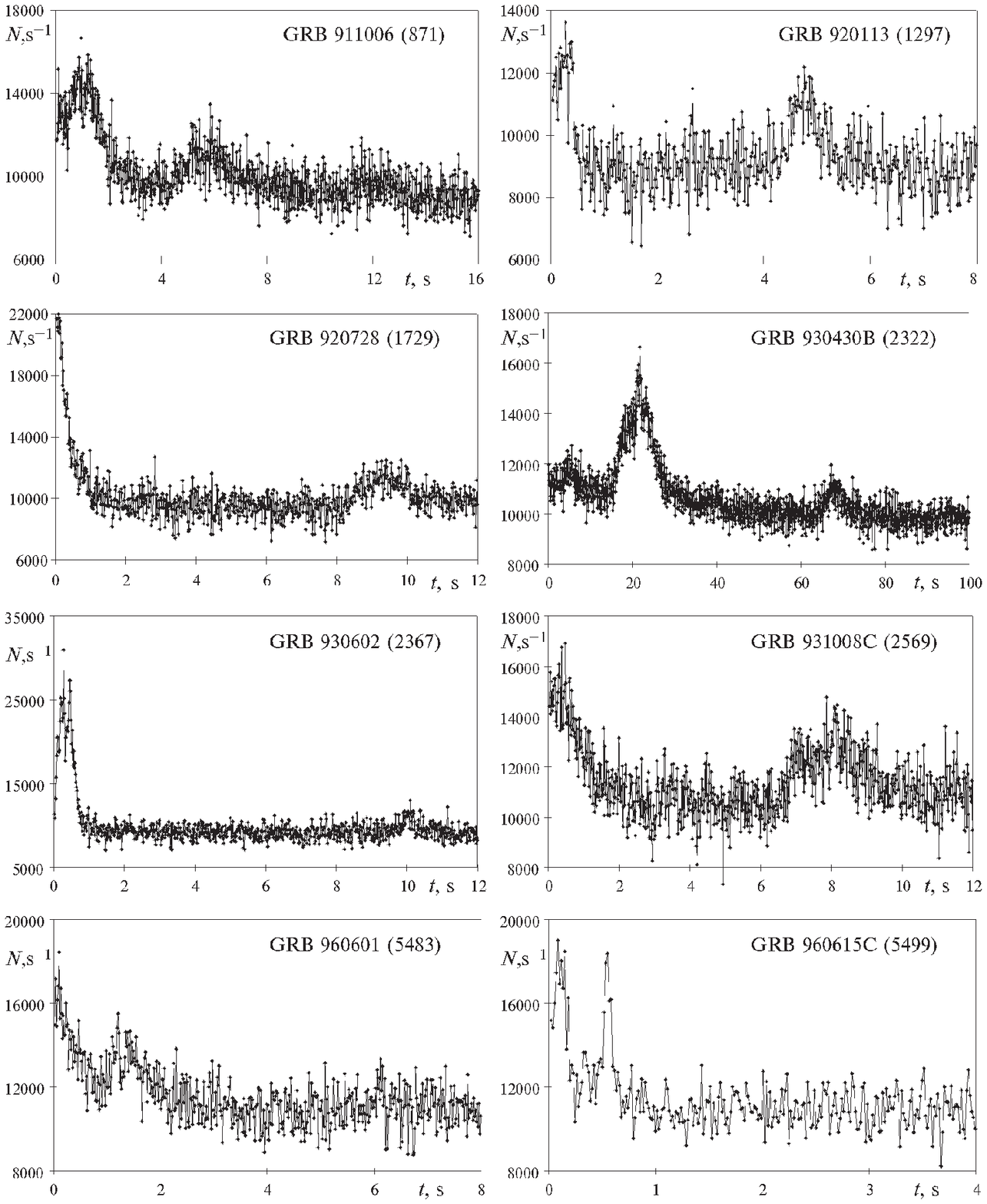}}
\vskip 50pt
\end{figure}

\begin{figure}[!h]
\centering
\resizebox{1.0\textwidth}{!}{\includegraphics{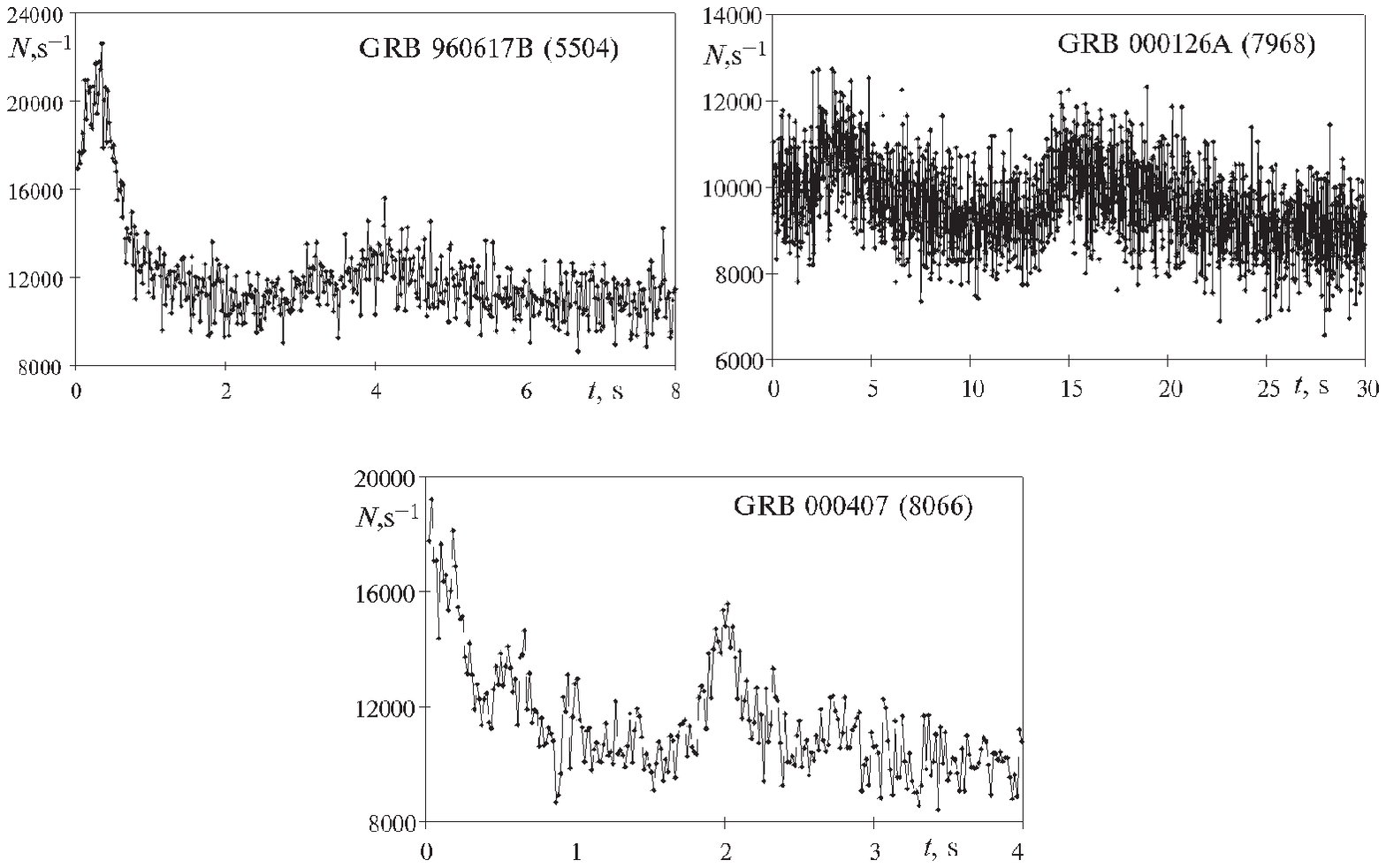}}
\caption{Lightcurves of possibly mesolensed gamma-ray bursts.}
\end{figure}

\subsection{Triple gamma-ray bursts}

The procedure of statistic test for component similarity for triple bursts
was the same as for double ones, best $\chi^2$-agreement method considered
all three components data at once. As the result of these tests we had not
found new candidates to mesolensed gamma-ray bursts, but third
weak component was found and satisfied all similar tests for two
double gamma-ray bursts listed above: GRB 911006 and GRB 930430B. As it
can be seen on Fig.3, third weak component follows two others for first of
these bursts and followed by them for second one. The parameters of relative
brightness $j_i$ and time delays $t_i-t_{i-1}$ for these bursts are given in
Table 2. Here parameters $j_i$ are normalized that brightness of the strongest
component is equal to unity. These values for two brightest components
slightly differ from that given in Table 1 for these two bursts due to third
component taking into account in $\chi^2$-tests.

\begin{table}[h]
\begin{center}
\caption{Triple bursts parameters.}
\begin{tabular}{|c|c|r|}
\hline
GRB & 911006 & 930430B \\
\hline
$j_1$          &  1.000 &  0.221 \\
$j_2$          &  0.404 &  1.000 \\
$j_3$          &  0.123 &  0.191 \\
$t_2-t_1, sec$ &  4.736 & 17.088 \\
$t_3-t_2, sec$ &  6.144 & 45.952 \\
$\Xi$          &  0.771 &  0.372 \\
\hline
\end{tabular}
\end{center}
\end{table}

Based on the values found we can try to calculate parameters of possible
near mesolensing in equations (\ref{11}) and (\ref{12}), that is $g$, $y$ and 
$x_t$. Since we don't know initial burst energy, brightness data for three
components give us only two parameters for this task, for example, normalized
brightness values for two weakest components $j_i$ given in table 2. Further,
since we don't know both $t_0$ and $C_0$ in equation (\ref{15}), time arrival
data give us just one parameter for lensing conditions calculation:
\begin{equation}
\label{17}
\Xi=\frac{t_2-t_1}{t_3-t_2}=\frac{T(x_2)-T(x_1)}{T(x_3)-T(x_2)}\mbox{.}
\end{equation}\\
The value of this parameter for two triple gamma-ray bursts is shown in the
last row of Table 2. Finally we have three independent parameters taken from
the observations and may search three corresponding mesolensing parameters.

The search was conducted for values of $x_t$ from $\sqrt{10}=3.16$ (that is
lower than for any known globular cluster in our Galaxy [12]) to infinity,
$g>1$, $0<|y|<y_c(g,x_t)$, where three lensed images are possible. This
procedure was made for both gamma-ray bursts GRB 911006 and GRB 930430B
and in both cases no corresponding lensing parameters were found.

This fact itself does not deny the possibility of near mesolensing of two 
triple gamma-ray bursts, since the lens density distribution may differ
from predicted one by King law. Moreover, brightness of one or more components
could be increased by microlensing effect at one singular star in a globular 
cluster, possibility of this effect being noticed in [10]. We may consider 
these two bursts to be the main candidates for near mesolensing, especially 
GRB 911006 with three components registered in decreasing order of intensity,
which is most probable in the mesolensing case.

Here we should add that the same gamma-ray burst GRB 911006 was localized in
thin annulus on the sky by triangulation measurements carried out by CGRO and 
``Ulysses" spacecraft [18] and bright ($13^m$) galaxy NGC 641 appeared to be 
situated in about $10^{'}$ from this annulus near its closest point to the 
BATSE error box center. Although annulus half-width by $3\sigma$-level is, as
paper [18] reports it two times less, about $5^{'}$.

\section{Discussion and conclusion}

In this work the possible gravitational mesolensing of cosmic gamma-ray
bursts was considered. The results of the search for this type of lensing
were more successful than that for macrolensing where no candidates were found.
All double gamma-ray bursts with similar lightcurve and spectra of components
had strong first and weak second pick on the lightcurve. It is this situation 
that is expected to be seen in the case of far mesolensing. The same picture
is observed for one of two triple bursts found, GRB 911006. Taking into
account the presence of bright galaxy close to its needle-type error box, we
may call it the most interesting candidate considered in this paper.

Surely each of found double or triple bursts could have such lightcurve
and spectral features due to the other, not related with gravitational lensing
reasons. In connection with that the question of mesolensing probability is 
actual. This value will be very small if we assume total mass of mesolenses in
the Universe to be such as observed in our Galaxy (about $10^{-3}$). But total
mass of these objects may be sufficiently increased at large $z$ [9], the role
of mesolenses can be played by the massive black holes, which number is 
expected to be very large [7,8]. We should also note that black hole will be
displayed as far (point mass) mesolens independently of its distance, and this
can be possible explanation of large number of double bursts with far 
mesolensing properties.

As in the case of macrolensing, the effect of mesolensing may occur only at 
extragalactic masses and observation of such event for gamma-ray bursts can be
considered as very essential confirmation of extragalactic nature of these
objects.

This work was done with financial assistance of Russian Fund for Basic
Research, grant 01-02-06247. Author would like to thank V.G.~Kurt,
B.V.~Komberg, Yu.V.~Baryshev, Yu.L.~Bukhmastova and Ya.Yu.~Tikhomirova for
their help and remarks.

\end{document}